\documentclass[preprint,12pt]{elsarticle} 
\usepackage{graphicx}
\usepackage{amsmath,amssymb}
\usepackage{eepic}
\usepackage{amsmath,amssymb,amsthm}
\usepackage{rotating}
\usepackage[update,prepend]{epstopdf}
\usepackage{graphicx}
\usepackage{multicol}
\newtheorem{theorem}{Theorem}
\newtheorem{definition}{Definition}

\journal{arXiv}

\begin{document}
\begin{frontmatter}
		
    \title{Estimation of Lyapunov dimension for the Chen and Lu systems}

    \author[spbu]{G. A. Leonov}

    \author[spbu,fin]{N. V. Kuznetsov\corref{cor}}
    \ead{nkuznetsov239@gmail.com}

    \author[spbu]{N. A. Korzhemanova}

    \author[spbu]{D. V. Kusakin}

    \cortext[cor]{Corresponding author}

    \address[spbu]{Faculty of Mathematics and Mechanics,
    	St. Petersburg State University, 198504 Peterhof,
        St. Petersburg, Russia}

    \address[fin]{Department of Mathematical Information Technology,
    University of Jyv\"{a}skyl\"{a}, \\ 40014 Jyv\"{a}skyl\"{a}, Finland}

 \begin{abstract}
 Nowadays various estimates of Lyapunov dimension
 of Lorenz-like systems attractors \cite{Leonov-2012-PMM} are actively developed.
 Within the frame of this study the question arises whether it is possible 
 to obtain the corresponding estimates of dimension for the Chen and Lu systems 
 using the reduction of them to the generalized Lorenz system \cite{Leonov-2013-DAN-ChenLu,LeonovK-2015-AMC}.
 In the work \cite{ChenY-2013} Leonov's method was applied for the estimation of Lyapunov dimension,
 and as a consequence the Lyapunov dimension of attractors of the Chen and Lu systems
 with the classical parameters was estimated.
 
 In the present work an inaccuracy in \cite{ChenY-2013} is corrected
 and it is shown that the revised domain of parameters, 
 where the estimate of Lyapunov dimension is valid, 
 does not involve the classical parameters of the Chen and Lu systems.
 \end{abstract}

 \begin{keyword}
 Lorenz-like system, Lorenz system, Chen system, Lu system,
 Kaplan-Yorke dimension, Lyapunov dimension, Lyapunov exponents.
 \end{keyword}

\end{frontmatter}

\section{Introduction}
 Nowadays various estimates of Lyapunov dimension of attractors of 
 generalized Lorenz systems \cite{Leonov-2012-PMM} 
\begin{equation}\label{sys-Lorenz-theorem}
 \left\{
 \begin{aligned}
 &\dot x= a(y-x)\\
 &\dot y= c x+d y-xz\\
 &\dot z=-b z+xy.
 \end{aligned}
 \right.
\end{equation}
are actively developed.
In the work  \cite{ChenY-2013} the following result
on the estimate of Lyapunov dimension of the generalized Lorenz system
 \eqref{sys-Lorenz-theorem} is formulated.

\begin{theorem}[\cite{ChenY-2013}]\label{ErrorTheorem}
 Let $K$ be invariant compact set of the generalized Lorenz system
 \eqref{sys-Lorenz-theorem}.
 Suppose that $a > 0$, $b >0$, $c+d >0$, $(a-b)(b+d)+ac>0$, and
 $q = \min(b, d)$.

 Then if it is satisfied one of the following conditions:
 \begin{enumerate}
 \item $d < 0$, $a^2 c \left(4 - \frac{b}{q}\right)
 - \frac{a^2 b(d+q)}{q} + 2a(b+d)(2a-3b) - b(b+d)^2 >0,$
 \item $d >0 $, $a > d,$
 \item d = 0;
 and also one of the conditions:
 \begin{enumerate}
 \item $2a \ge b$, $a^2 (3b + c) - b^3 - ab(6b+c) >0,$
 \item $2a < b$, $3a^2 (b+c) - 6 ab^2 - b^3 >0;$
 \end{enumerate}
 \end{enumerate}
 then
 $\dim_L K \le 3 - \frac{2 (a + b - d)}{a - d + \sqrt{(a+d)^2 + 4 ac}}$
 is valid.
\end{theorem}
However in the proof of this result in \cite{ChenY-2013}
there is an inaccuracy, the consideration of which the present work is devoted.

\section{Lyapunov dimension for the generalized Lorenz system}
Introduce a notion of Lyapunov dimension. Consider, for this purpose,
the linearization of dynamical system
\eqref{sys-Lorenz-theorem}
along a solution ${\bf \rm x}(t, {\bf \rm x}_0)$
\begin{equation}\label{eq:lin_eq}
 \frac{d{\bf \rm u}}{d t} = J({\bf \rm x}(t, {\bf \rm x}_0)) \, {\bf \rm u},
\end{equation}
where
$J({\bf \rm x}(t, {\bf \rm x}_0))$
is the $(n \times n)$ Jacobian matrix evaluated along the trajectory
 ${{\bf \rm x}(t, {\bf \rm x}_0)}$ of system \eqref{sys-Lorenz-theorem}.
The fundamental matrix $X(t,{\bf \rm x}_0) $ of linearized system \eqref{eq:lin_eq}
is defined by the variational equation
\begin{equation}\label{eq:var_eq}
 \dot{X}(t,{\bf \rm x}_0) =
 J({\bf \rm x}(t, {\bf \rm x}_0)) \, X(t,{\bf \rm x}_0).
\end{equation}

Let $\sigma_1 (X(t,{\bf \rm x}_0)) \geq \cdots
\geq \sigma_n (X(t,{\bf \rm x}_0)) > 0$ be singular values of
a fundamental matrix $X(t,{\bf \rm x}_0)$
(the square roots of the eigenvalues of the matrix
$X(t,{\bf \rm x}_0)^{*}X(t,{\bf \rm x}_0)$ are ordered for each $t$).
\begin{definition}\label{def:le}
 The {\it Lyapunov exponents (LEs)}\footnote{
 Two well-known definitions of Lyapunov exponents, which are used,
 are the following:
 the upper bounds of the exponential growth rate
 of the norms of linearized system solutions (LCEs) \cite{Lyapunov-1892}
 and the upper bounds of the
 exponential growth rate of the singular values
 of fundamental matrix of linearized system (LEs) \cite{Oseledec-1968}.
 These definitions usually give the same values,
 but for a particular system, LCEs and LEs may be different
 (see, e.g., corresponding example in \cite{KuznetsovAL-2014-arXiv}).
 In general, it may be shown that the ordered LCEs majorise
 the ordered LEs, thus the values of Kaplan-Yorke (Lyapunov) dimension based
 on LEs and LCEs may be different.
 See, e.g., a more detailed discussion
 in \cite{KuznetsovAL-2014-arXiv,LeonovK-2007}.

 Note, that LEs are independent of the choice \cite{KuznetsovAL-2014-arXiv}
 of fundamental matrix at the point ${\bf \rm x}_0$
 unlike the Lyapunov characteristic exponents
  (to get all possible values of LCEs, a
 \emph{normal fundamental matrix} has to be considered \cite{Lyapunov-1892}).
 }
 in the point ${\bf \rm x}_0$ are the numbers  (or symbols $\pm \infty$)
\begin{equation}\label{defLE}
 {\rm LE}_i ({\bf \rm x}_0) = \limsup_{t \to \infty}
 \frac{1}{t} \ln \sigma_i (X(t,{\bf \rm x}_0)).
\end{equation}
\end{definition}

Following \cite{KaplanY-1979}, introduce a notion of Lyapunov dimension.
\begin{definition}
 A local Lyapunov dimension of a point ${\bf \rm x}_0$
 in the phase space of dynamical system,
 generated by equation \eqref{sys-Lorenz-theorem}, is a number
 \begin{gather}\label{formula:kaplan}
 \dim_L {\bf \rm x}_0 = j({\bf \rm x}_0) +
 \cfrac{{\rm LE}_1({\bf \rm x}_0) + \ldots +
 {\rm LE}_j({\bf \rm x}_0)}{|{\rm LE}_{j+1}({\bf \rm x}_0)|},
 \end{gather}
 where ${\rm LE}_1({\bf \rm x}_0) \geq \ldots \geq {\rm LE}_j({\bf \rm x}_0),$
 are Lyapunov exponents;
 $j({\bf \rm x}_0) \in [1, n]$ is the smallest natural number $m$ such that
 $$
 {\rm LE}_1({\bf \rm x}_0) + \ldots + {\rm LE}_{m+1}({\bf \rm x}_0) < 0,
 \quad 
 \cfrac{{\rm LE}_1({\bf \rm x}_0) + \ldots +
 {\rm LE}_m({\bf \rm x}_0)}{|{\rm LE}_{m+1}({\bf \rm x}_0)|} < 1.
 $$
\end{definition}

The Lyapunov dimension of invariant set $K$ of dynamical system is defined by
\begin{equation}
 \dim_L K = \sup_{x_0 \in K} \dim_Lx_0.
\end{equation}

Note that LEs and Lyapunov dimension are invariant under the linear
change of variable (see, e.g., \cite{KuznetsovAL-2014-arXiv}).

Alongside with widely used numerical methods for estimating and computing
the Lyapunov dimension there exists analytical approach suggested by G.Leonov
\cite{Leonov-1991-Vest,LeonovB-1992,Leonov-2001,BoichenkoLR-2005,Leonov-2012-PMM}
that is based on the direct Lyapunov method.

Let us outline the main theorems describing this approach .
Let
$\lambda_1 ({\bf \rm x},S) \geqslant \cdots \geqslant \lambda_n ({\bf \rm x},S)$
be the eigenvalues of the following matrix
\begin{equation*}
 \frac{1}{2} \left( S J({\bf \rm x}) S^{-1} +
 (S J({\bf \rm x}) S^{-1})^{*}\right).
\end{equation*}

\begin{theorem}[\cite{Leonov-2002,Leonov-2012-PMM}]\label{theorem:th1}
Given an integer $j \in [1,n]$ and $s \in [0,1]$,
suppose that there is a continuously differentiable scalar function
$\vartheta: \mathbb{R}^n \rightarrow \mathbb{R}$
and a nonsingular matrix $S$ such that
\begin{equation}\label{ineq:th1}
 \lambda_1 ({\bf \rm x},S) + \cdots + \lambda_j ({\bf \rm x},S) + s\lambda_{j+1}
 ({\bf \rm x},S) + \dot{\vartheta}({\bf \rm x}) < 0,
 ~ \forall \, {\bf \rm x} \in K.
\end{equation}
Then $\dim_L K \leqslant j+s$.
\end{theorem}
Here $\dot{\vartheta}$ is derivative of the function
$\vartheta$ with respect to the system \eqref{sys-Lorenz-theorem}.

\begin{theorem}[\cite{Leonov-1991,LeonovB-1992,BoichenkoLR-2005,Leonov-2012-PMM}]
\label{theorem:th2}
Assume that there is a continuously differentiable scalar function $\vartheta$
and a nonsingular matrix $S$ such that
\begin{equation}\label{ineq:th2}
 \lambda_1 ({\bf \rm x},S) + \lambda_2 ({\bf \rm x},S)
 + \dot{\vartheta}({\bf \rm x}) < 0, ~
 \forall \, {\bf \rm x} \in \mathbb{R}^n.
\end{equation}
Then any solution of system \eqref{sys-Lorenz-theorem},
bounded on $[0,+\infty)$,
tends to an equilibrium as $t \rightarrow +\infty$.
\end{theorem}

Thus, if condition~\eqref{ineq:th2} holds,
then the global attractor of system \eqref{sys-Lorenz-theorem}
coincides with its stationary set.
For proving Theorem \ref{ErrorTheorem} in \cite{ChenY-2013}
it is used above described approach with
\begin{equation}\label{matrixSdefinition}
 S = \begin{pmatrix}
 \rho^{-1} & 0 & 0\\
 -\frac{b+d}{a} & 1 & 0\\
 0 & 0 & 1
 \end{pmatrix},
\end{equation}
where
\begin{equation}\label{rhodefinition}
 \rho=\frac{a}{\sqrt{(a-b)(b+d)+ac}},
\end{equation}
and the function
\begin{equation}
 \vartheta (x, y, z) = \frac{(1-s) V(x, y, z)}{2\sqrt{(a-2b -d)^2
 + \frac{4a^2}{\rho^2}}} = \frac{(1-s) V(x, y, z)}{2\sqrt{(a+d)^2 + 4ac}},
\end{equation}
where
\begin{equation}
 V(x, y, z) = \gamma_1\left(\frac{x^4}{4}-ax^2z\right)
 + \gamma_2 y^2 + (a^2 \gamma_1 + \gamma_2)z^2 + \gamma_3 xy
 + \gamma_4 x^2 - 4\frac{a}{b}z.
\end{equation}
The parameters $\gamma_1, \gamma_2, \gamma_3, \gamma_4$
are chosen in dependence of system parameters in such a way that
the conditions of Theorem \ref{theorem:th1} or \ref{theorem:th2} are satisfied.

 In proving Theorem \ref{ErrorTheorem} in the work  \cite{ChenY-2013}
 there is committed an inaccuracy, which does not permit one to obtain
 a condition for estimating the attractor dimension in the case $d >0$.
 Namely, in the proof there is an inaccurate assertion
 that in the case $2a + b -2d >0$ the inequality
\begin{equation}\label{cond1}
 \gamma_2 > \frac{\rho^2}{2b} - \left(\frac{a(2a + b - 2d)^2}{8b}
 + a^2\right)\gamma_1
\end{equation}
yields the inequality
\begin{equation}\label{cond2}
 \gamma_3 ^{(1)} \le N \gamma_1\le \gamma_3 ^{(2)},
\end{equation}
where $\gamma_1 > 0, N < 2ad, \gamma_3 ^{(1),(2)} = (2a^2 +ab)\gamma_1 \mp
\frac{\sqrt{-16a \gamma_1 (\rho^2 - 2a^2 b \gamma_1 - 2b \gamma_2)}}{2}$.
\medskip

Let us show that this assertion may be not true.
Consider specific values of system parameters
and running parameters for which the assertion is not satisfied.

Take $a = 1$, $b = 2$, $c = 6$, $d = 1$ for which $2a + b -2d=2$,
and $\gamma_1 = 5$, $\gamma_2 = -\frac{59}{12}$.
Then one has $N < 2$.

Condition \eqref{cond1} is satisfied since $-\frac{59}{12} > -\frac{37}{6}$.
However $\gamma_3^{(1)} = 20$ and consequently for $N<2$ the relation
$20 < 5N$ is impossible, i.e. condition \eqref{cond2} under such parameters
is not satisfied.

Correcting the above-mentioned inaccuracy, we obtain the following result.
\begin{theorem}\label{our}
 Denote by $K$ the bounded invariant set of system \eqref{sys-Lorenz-theorem},
 involving the point $x = y = z = 0$.
 Suppose that $a>0, b>0, d>0$, $c+d>0$, $ ac + (a - b) (b + d) > 0$,
  and there exists $\gamma $ such that the inequality  $$-d(2a+b+\gamma )^2 + 8a bd + 4\gamma a b >0$$
  is satisfied.

 Let $$4b(-d(2a +b+\gamma )^2 + 8a bd + 4\gamma a b)\left( \frac{\rho^2}{4b}
 + \frac{\rho^2(a+b+d)^2}{4a^2 (c+d)} - \frac{a}{b(c+d)}\right) +$$
 $$ + \rho^2 (b+d) (2a-b+\gamma)^2 < 0,$$
 where $\rho$ is defined as in \eqref{rhodefinition}.

 In this case
 \begin{enumerate}
 \item if
 \begin{equation}\label{r1}
 \left\{ \begin{gathered}
 a-d+2b > 0, \hfill \\
 c < (b -d)\left(\frac{b}{a} + 1\right), \\
 \end{gathered} \right.
 \end{equation}
 then any bounded on $[0; +\infty)$ solution of system \eqref{sys-Lorenz-theorem}
 tends to a certain equilibrium as $t \to +\infty$;

 \item if
 \begin{equation}\label{r2}
 \left\{ \begin{gathered}
 a-d+b\ge0, \hfill \\
 c > (b -d)\left(\frac{b}{a}+1\right), \\
 \end{gathered} \right.
 \end{equation}
 then
 \begin{equation}\label{formula}
 \dim_L K = 3 - \frac{2 (a + b -d)}{a -d + \sqrt{(a+d)^2 + 4 ac}}
 \end{equation}
 \end{enumerate}
\end{theorem}

For proving Theorem \ref{our} we uses the matrix $S$,
defined according to \eqref{matrixSdefinition}, and the function
\begin{equation}\label{theta}
 \vartheta (x, y, z) = \frac{(1-s) V(x, y, z)}{[(a+d)^2+4ac]^{\frac{1}{2}}},
\end{equation}
where
\begin{equation}\label{V}
 V(x, y, z) = \gamma_4 x^2+(-a\gamma_1+\gamma_3)y^2+\gamma_3z^2+
 \frac{1}{4a}\gamma_1 x^4-\gamma_1 x^2 z - \gamma_1 \gamma_2 xy -\frac{a}{b}z.
\end{equation}
 Exact value of dimension is obtained here
 from the comparison of dimension estimate \eqref{ineq:th2}
 and the value of Lyapunov dimension in zero point.
 The obtained value corresponds to the dimension of global attractor
 or a (local) B-attractor, involving zero equilibrium.


\section{Lyapunov dimension for the Chen and Lu systems}

In 2012 G.A.~Leonov suggested to consider
for the Chen and Lu systems the following change of variables
\cite{Leonov-2013-DAN-ChenLu}
\begin{equation}\label{trans-LEO}
 x \rightarrow hx,\ y \rightarrow hy,\ z \rightarrow hz,\ t \rightarrow h^{-1}t
\end{equation}
with $h=a$.
Under this transformation for $a\neq 0$ the Chen and Lu systems
are transformed to the generalized Lorenz system \eqref{sys-Lorenz-theorem}
with the parameters
\begin{equation}\label{newparameters}
 a \to 1 , b \to \frac{b}{a}, d \to \frac{d}{a}.
\end{equation}
Note that for $a=0$ the Chen and the Lu systems become linear
and their dynamics have minor interest.
Thus, without loss of generality, one can assume that $a=1$.
The transformation \eqref{trans-LEO}
with $h=a$ does not change the direction of time
for the positive chaotic parameters considered
in the works \cite{ChenU-1999,LuChen-2002}.
Remark that the applying of transformations with the time inversion
(see corresponding discussion of
the Lorenz, Chen and Lu systems \cite{LeonovK-2015-AMC})
is not suitable for the study of Lyapunov dimension
of invariant sets
since the absolute values of Lyapunov exponents and corresponding
local Lyapunov dimension
in direct and backward time may be different
(see, e.g., \cite{LeonovK-2007,GelfertM-2010,LeonovK-2014-arXiv-LCL,LeonovK-2015-AMC}).

Below there are represented domains in the obtained
two-dimensional parameter space \eqref{newparameters}
for which the condition of Theorem \ref{our} is satisfied.
On a plane there are also shown the points, corresponding
to standard parameters of chaotic attractors.
For the Lu system such parameters are $(a, b, d) = (36, 3, 20)$,
and after the change $(b, d)=(\frac{1}{12}, \frac{5}{9})$;
for the Chen system --- $(a, b, d) = (35, 3, 28)$
and $(b, d) = (\frac{3}{35}, \frac{4}{5})$.

\begin{figure}[t]
 \begin{multicols}{2}
 \hfill
 \includegraphics[width=0.75\linewidth, angle=270]{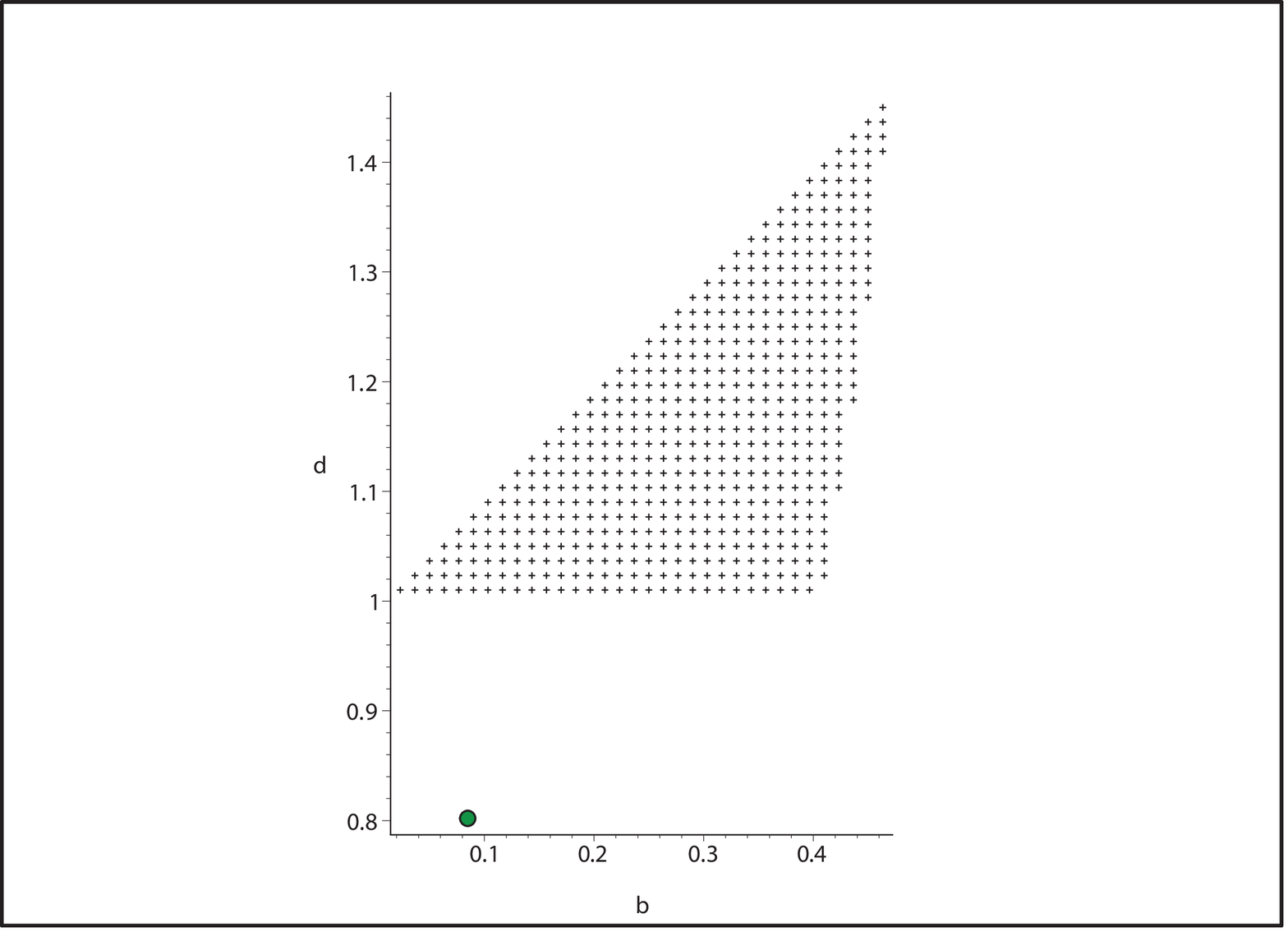}
 \hfill
 \caption{Two-parametric domain for the Chen system}
 \label{figLeft}
 \hfill
 \includegraphics[width=0.75\linewidth, angle=270]{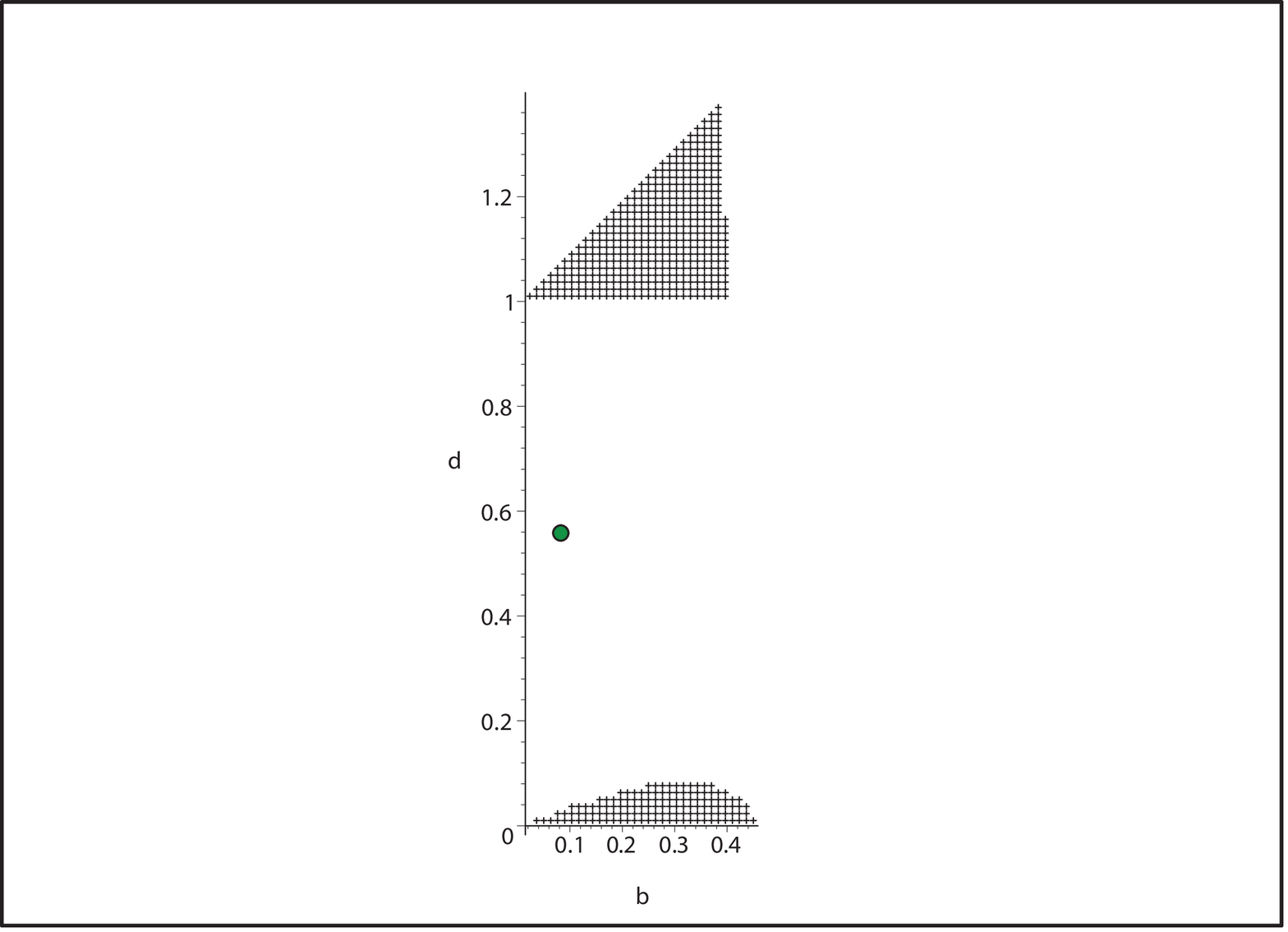}
 \hfill
 \caption{Two-parametric domain for the Lu system}
 \label{figRight}
 \end{multicols}
\end{figure}

 As shown in Figs \ref{figLeft}-\ref{figRight},
 nor the Lu system nor the Chen system standard parameters
 are situated in the obtained domain.
 While Lyapunov dimension of the classical Lorenz attractor has been obtained analytically,
 for the Chen and Lu attractors it is still an open problem.
 Also by numerical simulation we do not find any chaotic attractor of the Chen and Lu systems with the parameters
 from the obtained domains.

Applying the same approach, we extend the results,
obtained in the case $d = 0$ in Theorem \ref{ErrorTheorem}.
\begin{theorem}\label{YangTheorem}
 Suppose that $a > 0, b >0, c > 0$, and the inequality
 $ ac + b(a - b) > 0$ is satisfied.

 Let $\gamma ^{(II)} > 0$ be valid, where $\gamma ^{(II)}$
 is the greater root of the equation
 \begin{equation}\label{YangEquation}
 \rho^2 (2a - b + \gamma)^2 + 16 a b \gamma
 \left(\frac{\rho^2}{4b}
 + \frac{\rho^2 (a + b)^2}{4 a ^2 c}
 - \frac{a}{bc}\right) = 0,
 \end{equation}
 where $\rho$ is given according to \eqref{rhodefinition} for $d=0$.

 In this case
 \begin{enumerate}
 \item if
 \begin{equation}
 c < b\left(\frac{b}{a} + 1\right),
 \end{equation}
 then any bounded on $[0; +\infty)$ solution of system
 \eqref{sys-Lorenz-theorem} tends to a certain equilibrium
 as $t \to +\infty$;

 \item if
 \begin{equation}
 c > b\left(\frac{b}{a}+1\right),
 \end{equation}
 then
 \begin{equation}
 dim_L K = 3 - \frac{2 (a + b)}{a + \sqrt{{a}^2 + 4 ac}}.
 \end{equation}
 \end{enumerate}
\end{theorem}
 Consider, for example, the parameters $a=1, b = 0.8, c =9.2$
for which there exists a chaotic attractor.
For them $2a =2$, i.e. $2a \ge b$ is satisfied.
However $a^2 (3b + c) - b^3 - ab(6b+c) = -0.112 < 0$. Thus, for the
above-mentioned parameters Theorem \ref{ErrorTheorem} cannot be applied.
But for them Theorem \ref{YangTheorem} can be applied
since for them a higher root of equation \eqref{YangEquation}
is equal to $\gamma^{(II)} \approx 8.583541432$.
\begin{figure}[t]
 \centering
 \includegraphics[width=0.5\linewidth, angle=270]{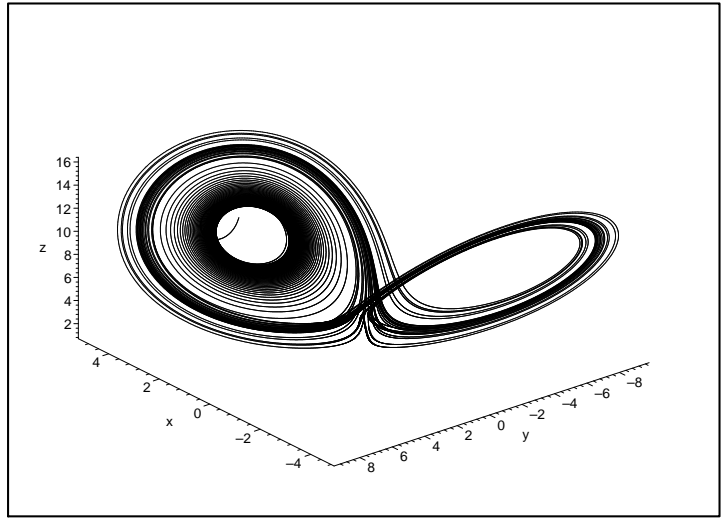}
 \hfill
 \caption{a = 1, b = 0.8, c = 9.2, d=0, $(x_0, y_0, z_0) = (5,0,7)$}
\end{figure}

Similarly it is possible to consider the generalized Lorenz system \eqref{sys-Lorenz-theorem}
not only with positive $a, b$, but with the negative ones.
\begin{theorem}\label{Main}
Suppose that $K$ is a bounded invariant set of system
 \begin{equation}\label{sys-Lorenz}
 \left\{
 \begin{aligned}
 &\dot x= \sigma(y-x)\\
 &\dot y= r x-d y-xz\\
 &\dot z=-b z+xy,
 \end{aligned}
 \right.
 \end{equation}
 involving the point $x = y = z = 0$.
 Suppose that $\sigma , b \ne 0$, the inequality
 $ r \sigma + (\sigma - b) (b - d) > 0$ is valid,
 and there exist numbers $\gamma_1, \gamma_2, \gamma_3$ such that
 \begin{equation}
 \begin{aligned}\label{first_condition}
 & \gamma_1 \ge 0,\\
 \end{aligned}
 \end{equation}
 \begin{equation}
 \begin{aligned}\label{second_condition}
 & 2d\gamma_3 > \frac{\rho^2}{4}+2d\sigma \gamma_1
 -\sigma \gamma_1 \gamma_2,\\
 \end{aligned}
 \end{equation}
 \begin{equation}
 \begin{aligned}\label{third_condition}
 & 2b\gamma_3 \ge \frac{\rho^2}{4} +
 \frac{\gamma_1}{4}(2\sigma+b+\gamma_2)^2,\\
 \end{aligned}
 \end{equation}
 \begin{equation}
 \begin{aligned}\label{fourth_condition}
 & 2(r-d)\gamma_3 \le 2\sigma\gamma_1(r-d)+\gamma_1\gamma_2(r-d)
 -\frac{\rho^2(b+\sigma-d)^2}{4\sigma^2}+\frac{\sigma}{b},\\
 \end{aligned}
 \end{equation}
 where $\rho = \frac{\sigma}{\sqrt{\sigma r + (\sigma-b)(b-d)}}$.

 In this case
 \begin{enumerate}
 \item if
 \begin{equation}
 \left\{
 \begin{gathered}
 \sigma+d+2b > 0, \hfill \\
 \sigma r < b^2 + \sigma d + \sigma b + bd,
 \end{gathered}
 \right.
 \end{equation}
 then any bounded on $[0; +\infty)$ solution of system
 \eqref{sys-Lorenz}  tends
 to a certain equilibrium as $t \to +\infty$;
 \item if
 \begin{equation}
 \left\{
 \begin{gathered}
 \sigma+d+b > 0, \hfill \\
 \left[
 \begin{gathered}
 \sigma + d >0, \hfill \\
 \left\{
 \begin{gathered}
 \sigma + d \le 0, \hfill \\
 \sigma(r - d) >0, \\
 \end{gathered}
 \right.\hfill
 \end{gathered}
 \right. \hfill \\
 \left[
 \begin{gathered}
 \sigma + d + 2b < 0, \hfill \\
 \left\{
 \begin{gathered}
 \sigma + d+ 2b \ge 0, \hfill \\
 \sigma r > b^2 + \sigma d + \sigma b + bd, \\
 \end{gathered}
 \right.
 \end{gathered}
 \right.
 \end{gathered}
 \right.
 \end{equation}
 then
 \begin{equation}
 dim_L K = 3 - \frac{2 (\sigma + b + d)}{\sigma + d
 + \sqrt{(\sigma-d)^2 + 4 \sigma r}}
 \end{equation}
 \end{enumerate}

\end{theorem}

\section*{References}
\bibliographystyle{elsarticle-num}

\end{document}